\documentclass[a4paper]{iopart}

\usepackage{graphicx}

\newcommand{\aver}[1]{{\left\langle\vphantom{\big|}#1\right\rangle}}
\newcommand{\smaver}[1]{{\langle #1\rangle}}
\newcommand{\D}{\mathalpha{\mathrm{d}}}
\newcommand{\I}{\mathalpha{\mathrm{i}}}
\newcommand{\Exp}[1]{\mathalpha{\mathrm{e}}^{\mbox{\footnotesize$#1$}}}

\renewcommand{\submitto}[1]{\relax}
\renewcommand{\maketitle}{\relax}
\makeatletter
\renewcommand{\ps@myheadings}{%
\renewcommand{\@oddhead}{\@empty}
\renewcommand{\@evenhead}{\@empty}
\renewcommand{\@oddfoot}%
{\hfil\textrm{\scriptsize(Posted on the arXiv on 31 July 2009)}\hfil}
\renewcommand{\@evenfoot}{\@oddfoot}}
\makeatother
\begin{document}

\title{Average transmission probability of a random stack}

\author{Yin Lu,$^{1,2}$ Christian Miniatura$^{1,3,4}$
and Berthold-Georg Englert$^{1,4}$}

\address{$^1$Centre for Quantum Technologies, National University of
  Singapore, Singapore 117543, Singapore}
\address{$^2$NUS Graduate School of Integrative Sciences and Engineering,
  Singapore 117542, Singapore} 
\address{$^3$Institut Non Lin\'eaire de Nice, UMR 6618, UNS, CNRS; 
  06560 Valbonne, France}
\address{$^4$Department of Physics, National University of Singapore,
  Singapore 117542, Singapore} 

\ead{\mailto{loisluyin@gmail.com}, 
\mailto{cqtmc@nus.edu.sg}, \mailto{cqtebg@nus.edu.sg}}

\begin{abstract}
The transmission through a stack of identical slabs that are separated by
gaps with random widths is usually treated by calculating the average of the
logarithm of the transmission probability.
We show how to calculate the average of the transmission probability itself
with the aid of a recurrence relation and derive analytical upper and lower
bounds. 
The upper bound, when used as an approximation for the transmission
probability, is unreasonably good and we conjecture that it is asymptotically
exact.  
\end{abstract}


\submitto{EJP}

\maketitle

\section{Introduction}
\label{sec:intro}
\begin{figure}[t]
\begin{center}
\includegraphics{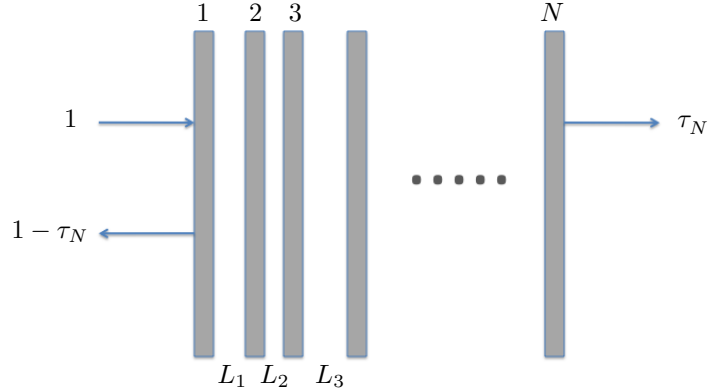}
\caption{A stack of $N$ identical slabs, each with single-slap transmission
  probability $\tau_1^{\ }$. The stack as a whole has transmission probability  
  $\tau_N^{\ }$, which depends on the phases that result from the random
  spacing of the slabs. 
  We are interested in $\smaver{\tau_N^{\ }}$, the transmission
  probability of the stack averaged over the ${N-1}$ phases.}
\label{fig:stack}
\end{center}
\end{figure}

We revisit a classical problem: The transmission through a linear array of many
identical slabs (glass plates, plastic transparencies, or the like)
with random separation, as depicted in Fig.~\ref{fig:stack}.
The transmission probability that Stokes derived in 1862 \cite{Stokes:62} 
on the basis of ray-optical arguments (thereby improving on an earlier attempt
by Fresnel in 1821; see Refs.~\cite{Buchwald:89} and \cite{Berry+1:97} for the
history of the subject) is not correct because
there are crucial interference effects that require a proper wave-optical
treatment.
Just that was given by Berry and Klein in 1997 \cite{Berry+1:97}
who found that the average of
the logarithm of the transmission probability through $N$ slabs is equal
to $N$ times the logarithm of the single-slab transmission probability,
\begin{equation}
  \label{eq:logaver}
  \aver{\log\tau_N^{\ }} = N \log\tau_1^{\ }\,.
\end{equation}
Here, $\tau_1^{\ }$ is the probability of transmission through a single slab and
$\tau_N^{\ }$ denotes the transmission probability for $N$ slabs. 
Its implicit dependence on the random phases that originate in the random
spacing of the slabs is averaged over, indicated by the $\smaver{\cdots}$
notation.    
As emphasized in Ref.~\cite{Berry+1:97}, the disorder is crucial; 
without it, most wavelength components would be
transmitted, and the stack should then appear rather
transparent, but this is not the case as a simple experiment with a stack of
transparencies demonstrates \cite{Berry+1:97,Hecht+1:74}. 

It is indeed common to average logarithms because they are known to be ``self
averaging'' \cite{Anderson+3:80}, and the exact result (\ref{eq:logaver}) is
truly remarkable.
But one should realize what it tells us about the average
transmission probability $\aver{\tau_N^{\ }}$ itself. 
As a consequence of the inequality
\begin{equation}
  \label{eq:logineq}
  \aver{\log\tau_N^{\ }}\leq\log\aver{\tau_N^{\ }}\,
\end{equation}
the Berry--Klein relation (\ref{eq:logaver}) amounts
to a lower bound on the average transmission probability,
\begin{equation}
  \label{eq:BK-lowlim}
  \aver{\tau_N^{\ }}\geq\tau_1^N\,.
\end{equation}
As we shall see below, this bound is not particularly tight because there is a
very large range of individual $\tau_N^{\ }$ values.
In particular, we note that the ray-optics result~\cite{Berry+1:97}
\begin{equation}
  \label{eq:rayoptics}
   \aver{\tau_N^{\ }}_{\mathrm{ray}}^{\ }
  =\frac{\tau_1^{\ }}{\tau_1^{\ }+N(1-\tau_1^{\ })}
\end{equation}
is consistent with (\ref{eq:BK-lowlim}).

It is the objective of the present contribution to report good wave-optics
estimates for $\aver{\tau^{\ }_N}$ and closely related quantities.
In particular, we will improve on the lower bound of (\ref{eq:BK-lowlim}) and
supplement it with an upper bound.
We observe that the upper bound, when used as an approximation for
$\aver{\tau_N^{\ }}$, is unreasonably good and seems to give us the exact
asymptotic values of quantities such as 
$\aver{\tau_{N+1}^{\ }}\big/\aver{\tau_{N}^{\ }}$ or $\aver{\tau_N^{\ }}^{1/N}$. 
At present, this coincidence of the upper bound with exact asymptotic values 
is a poorly understood mystery.

\section{Single slab: The transfer matrix}
\label{sec:1slab}
\begin{figure}[t]
\begin{center}
\includegraphics{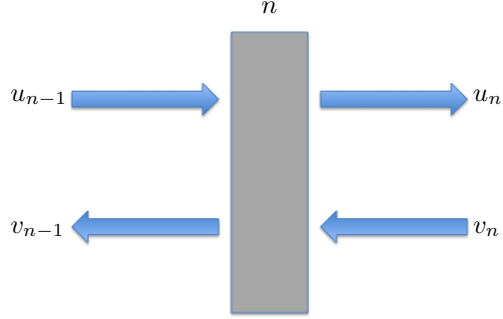}
\caption{\label{fig:slab}%
  Amplitudes on both sides of the $n$th slab. 
  The unitary scattering matrix $S$ of (\ref{eq:S}) relates the incoming
  amplitudes $u_{n-1}$ and $v_n$ to the outgoing amplitudes $u_n$ and
  $v_{n-1}$, whereas the transfer matrix $T$ of (\ref{eq:T}) connects the
  amplitudes on the left with the amplitudes on the right.}
\end{center}
\end{figure}
For a wave of wavelength $2\pi/k$, the wave functions  
to the left and to the right of the $n$th slab are 
\begin{eqnarray}
  \label{eq:wf-n}
  \psi_n^{\mathrm{(left)}}(x)&=&u_{n-1}\Exp{\I k (x-x_n)}
                              +v_{n-1}\Exp{-\I k (x-x_n)}\,,\nonumber\\
  \psi_n^{\mathrm{(right)}}(x)&=&u_{n}\Exp{\I k (x-x_n-\ell)}
                              +v_{n}\Exp{-\I k (x-x_n-\ell)}\,,
\end{eqnarray}
where $x_n$ is the position of the left edge and $\ell$ is the
thickness of the slab; see Fig.~\ref{fig:slab}.
The incoming amplitudes are related to the outgoing amplitudes by the unitary
\emph{scattering matrix} $S$,
\begin{equation}
  \label{eq:S}
  \left(\begin{array}{c} u_n \\ v_{n-1} \end{array}\right)=S
  \left(\begin{array}{c} u_{n-1} \\ v_n \end{array}\right)=
  \left(\begin{array}{cc} a & b' \\ b & a' \end{array}\right)
  \left(\begin{array}{c} u_{n-1} \\ v_n \end{array}\right),
\end{equation}
where the entries of $S$ are restricted by
\begin{equation}
  \label{eq:Sab}
  |a|^2=|a'|^2=\tau^{\ }_1\,,\quad |b|^2=|b'|^2=1-\tau^{\ }_1\,,
   \quad a^*b'+b^*a'=0\,,
\end{equation}
which account for the single-slab transmission probability~$\tau_1^{\ }$ and the
unitary nature of~$S$.  
The particular values of the complex phases of $a$, $b$, $a'$, and $b'$ are of
secondary interest, but we note that we have ${a=a'=\Exp{\I k\ell}}$ and
${b=b'=0}$ for a completely transparent, non-scattering slab, for which
${\psi_n^{\mathrm{(left)}}(x)=\psi_n^{\mathrm{(right)}}(x)}$.

The \emph{transfer matrix} $T$ is used to express the amplitudes on the right
in terms of the amplitudes on the left,
\begin{equation}
  \label{eq:T}
  \left(\begin{array}{c} u_n \\ v_n \end{array}\right)=T
  \left(\begin{array}{c} u_{n-1} \\ v_{n-1} \end{array}\right)\,.  
\end{equation}
The one-to-one relation between $S$ and $T$ implies that the transfer matrix
is of the form
\begin{equation}
  \label{eq:genT}
  T=\Exp{\I\alpha}
  \left(\begin{array}{cc}\Exp{\I\beta} & 0\\[1ex]
                         0 & \Exp{-\I\beta}\end{array}\right)
  \left(\begin{array}{cc} \cosh\theta & \sinh\theta \\[1ex] 
                          \sinh\theta & \cosh\theta \end{array}\right)
  \left(\begin{array}{cc}\Exp{\I\beta'} & 0\\
                         0 & \Exp{-\I\beta'}\end{array}\right),
\end{equation}
where 
\begin{equation}
  \label{eq:tau-theta}
    \cosh\theta=\frac{1}{\sqrt{\tau_1^{\ }}}\,,\quad 
    \tau_1^{\ }=\frac{2}{\cosh(2\theta)+1}\,,
\end{equation}
and $\alpha$, $\beta$, $\beta'$ are phase factors that have fixed values
which, however, are largely irrelevant for what follows.

The transfer matrix for the gap of length $L_n$ between the $n$th slab and the
$(n+1)$th slab is the diagonal phase matrix
\begin{equation}
  \label{eq:D}
  D(kL_n)=  \left(\begin{array}{cc}\Exp{\I kL_n} & 0\\[1ex]
                         0 & \Exp{-\I kL_n}\end{array}\right)\,.
\end{equation}
Phase matrices of the same structure sandwich the central $\theta$-dependent
matrix in (\ref{eq:genT}), so that we have
\begin{equation}
  \label{eq:genT'}
  T=\Exp{\I\alpha}D(\beta)\,t(\theta)\,D(\beta')\quad\mbox{with}\enskip
  t(\theta)=\left(\begin{array}{cc} \cosh\theta & \sinh\theta \\[1ex] 
                  \sinh\theta & \cosh\theta \end{array}\right)
\end{equation}
as a more useful way of writing $T$.

The product of two transfer matrices is another transfer matrix, whereby the
relevant observation is the composition law
\begin{equation}
  \label{eq:tDt}
  t(\theta_1)\,D(\phi)\,t(\theta_2)=D(\gamma)\, t(\theta_{1\&2})\,D(\gamma') 
\end{equation}
with $\theta_{1\&2}$ determined by
\begin{eqnarray}
  \label{eq:newtheta}
  \cosh(2\theta_{1\&2})&=&\cosh(2\theta_1)\cosh(2\theta_2)
\nonumber\\&&\mbox{}   +\cos(2\phi)\sinh(2\theta_1)\sinh(2\theta_2)
\end{eqnarray}
and the phases $\gamma$ and $\gamma'$ by
\begin{eqnarray}
  \label{eq:gammas}
  \Exp{2\I\gamma}\sinh(2\theta_{1\&2})&=&\sinh(2\theta_1)\cosh(2\theta_2)
\nonumber\\&&\mbox{}            +\cos(2\phi)\cosh(2\theta_1)\sinh(2\theta_2)
\nonumber\\&&\mbox{}            +\I\sin(2\phi)\sinh(2\theta_2)\,,
\nonumber\\
  \Exp{2\I\gamma'}\sinh(2\theta_{1\&2})&=&\cosh(2\theta_1)\sinh(2\theta_2)
\nonumber\\&&\mbox{}             +\cos(2\phi)\sinh(2\theta_1)\cosh(2\theta_2)
\nonumber\\&&\mbox{}             +\I\sin(2\phi)\sinh(2\theta_1)\,.
\end{eqnarray}
Whereas (\ref{eq:gammas}) is of no consequence for the following
considerations, the rule (\ref{eq:newtheta}) is of central importance.

\section{Many slabs: A recurrence relation}
\label{sec:Nslabs}
We now turn to the situation of Fig.~\ref{fig:stack}, where we have $N$
identical slabs separated by gaps $L_1$, $L_2$, \dots, $L_{N-1}$ that are not
controlled on the scale set by the wavelength $2\pi/k$.
Therefore, we regard the phase factors $\Exp{\I kL_n}$ as random with a
uniform distribution on the unit circle in the complex plane. 

The over-all transfer matrix
\begin{eqnarray}
  \label{eq:totT}
  T_{\mathrm{tot}}&=&TD(kL_1)TD(kL_2)T\cdots TD(kL_{N-1})T\nonumber\\
             &=&\Exp{\I N\alpha}D(\beta)
                \Biggl(\prod_{n=1}^{N-1}t(\theta)\,D(\phi_n)\Biggr)
                t(\theta)\,D(\beta')\nonumber\\
             &=&\Exp{\I\alpha_{\mathrm{tot}}}D(\beta_{\mathrm{tot}})
                \,t(\theta_{\mathrm{tot}})\,D(\beta'_{\mathrm{tot}})
\end{eqnarray}
is characterized by $\theta_{\mathrm{tot}}$ which is obtained by repeated
application of the composition rule (\ref{eq:newtheta}), whereby the phases
${\phi_n=\beta+\beta'+kL_n}$ have random values. 
Each experimental realization of the $N$-slab stack of Fig.~\ref{fig:stack}
has different values for these random phases, and the transmission probability
\begin{equation}
  \label{eq:tauN}
  \tau_N^{\ }=\frac{2}{\cosh(2\theta_{\mathrm{tot}})+1}
\end{equation}
varies from one experiment to the next.
We need to average over the ${N-1}$ random phases to find $\aver{\tau_N^{\ }}$.

\begin{figure}[t]
\begin{center}
\includegraphics{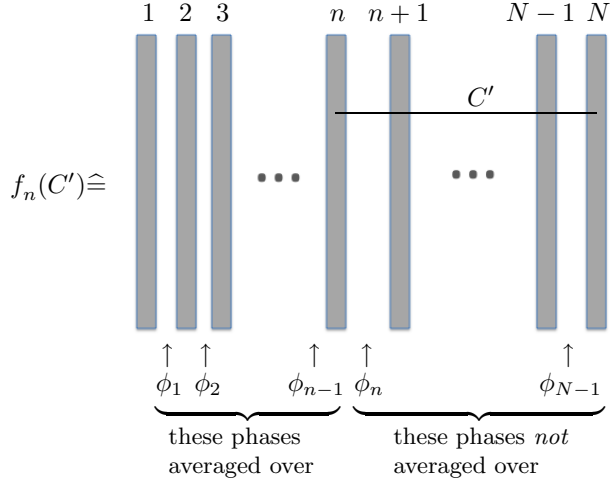}
\caption{\label{fig:recursion}%
   Regarding the meaning of $f_n^{\ }(C')$ in (\ref{eq:recurrence}) and
   (\ref{eq:faver}). 
   The random phases $\phi_1$,
   $\phi_2$, \dots, $\phi_{n-1}$ have already been averaged over, but the
   averaging over the phases $\phi_n$, \dots, $\phi_{N-1}$ is yet to be
   performed.  
}
\end{center}
\end{figure}

Let us consider a somewhat more general question: What is the average value
$\aver{f\bigl(\cosh(2\theta_{\mathrm{tot}})\bigl)}$ of a function of
$\cosh(2\theta_{\mathrm{tot}})$, and thus of a function of $\tau_N^{\ }$? 
When the averaging is carried out successively, first averaging over $\phi_1$,
then over $\phi_2$, and so forth, finally over $\phi_{N-1}$, we have
an intermediate value $f_n^{\ }(C')$ after averaging over the first ${n-1}$
random phases; see Fig.~\ref{fig:recursion}.
Here $C'$ denotes the value of $\cosh(2\theta_{n\cdots N})$ for the stack of
slabs $n$ to $N$ with its dependence on the remaining phases $\phi_n$, \dots,
$\phi_{N-1}$. 
We then have $f_1^{\ }(C')=f(C')$ for the value prior to any averaging, and
(\ref{eq:newtheta}) tells us that we get $f_{n+1}^{\ }(C')$ from $f_n^{\
}(C')$ by means of 
\begin{eqnarray}
  \label{eq:recurrence}
&&f_{n+1}^{\ }(C')=\int\limits_{(2\pi)}\frac{\D\varphi}{2\pi}\,
                   f_n^{\ }(CC'+SS'\cos\varphi)
\nonumber\\&&
\mbox{with}\enskip C=\cosh(2\theta)=\frac{2}{\tau_1^{\ }}-1\,,\quad
                   S=\sinh(2\theta)=\frac{2}{\tau_1^{\ }}\sqrt{1-\tau_1^{\ }}\,,
\nonumber\\&&
\mbox{and}\enskip S'=\sqrt{{C'}^2-1}\quad\mbox{with}\enskip C'\geq1\,,
\end{eqnarray}
and the integration covers any convenient interval of $2\pi$.
Eventually this takes us to 
\begin{equation}
  \label{eq:faver}
  \aver{f\bigl(\cosh(2\theta_{\mathrm{tot}})\bigl)}=f_N^{\ }(C)
\end{equation}
when the recursive averaging over the ${N-1}$ random phases is completed.

For illustration, we take $f_1^{\ }(C')=C'$ as a first example.
The recurrence relation (\ref{eq:recurrence}) yields $f_n^{\ }(C')=C^{n-1}C'$, 
so that 
\begin{equation}
  \label{eq:1stex}
   \aver{\cosh(2\theta_{\mathrm{tot}})}=\cosh(2\theta)^N
\end{equation}
or
\begin{equation}
  \label{eq:1stex'}
  \aver{1/\tau_N^{\ }}=\frac{1}{2}+\frac{1}{2}(2/\tau_1^{\ }-1)^N
\end{equation}
when stated in terms of transmission probabilities.
A second illustrating example is $f_1^{\ }(C')={C'}^2-\frac{1}{3}$, for which
\begin{eqnarray}
  \label{eq:2ndex}
  f_n^{\ }(C')&=& \biggl[\frac{3}{2}f_1^{\ }(C)\biggr]^{n-1}f_1^{\ }(C')\,,
\nonumber\\
  f_N^{\ }(C)&=&\frac{2}{3}\biggl[\frac{3}{2}f_1^{\ }(C)\biggr]^N\,,
\end{eqnarray}
and
\begin{equation}
  \label{eq:2ndex'}
   \aver{\cosh(2\theta_{\mathrm{tot}})^2}=\frac{1}{3}
    +\frac{2}{3}\biggl[\frac{1}{2}\bigl(3\cosh(2\theta)^2-1\bigr)\biggr]^N
\end{equation}
follows.

Taken together, (\ref{eq:1stex}) and (\ref{eq:2ndex'}) tell us that the
normalized variance of $\cosh(2\theta_{\mathrm{tot}})$ grows exponentially
with the number of slabs,
\begin{eqnarray}
  \label{eq:Cvar}
  \frac{\aver{\cosh(2\theta_{\mathrm{tot}})^2}}
       {\aver{\cosh(2\theta_{\mathrm{tot}})}^2}-1
  &=&\frac{1}{3}\cosh(2\theta)^{-2N}
   +\frac{2}{3}\biggl[1+\frac{1}{2}\tanh(2\theta)^2\biggr]^N\nonumber\\
  &\approx&\frac{2}{3}\biggl[\frac{3}{2}
         -\frac{1}{2}\Bigl(\frac{\tau_1^{\ }}{2-\tau_1^{\ }}\Bigr)^2\biggr]^N
   \quad\mbox{for ${N\gg1}$.}
\end{eqnarray}
The values of $\tau^{\ }_N$ cover a correspondingly large range, and so we
understand why the two sides of (\ref{eq:logineq}) differ by much.

This brings us to the much more important $\log\tau$ case of 
\begin{equation}
  \label{eq:3rdex}
  f_1^{\ }(C')=\log\frac{2}{C'+1}\,.
\end{equation}
Here,
\begin{equation}
  \label{eq:3rdex'}
  f_n^{\ }(C')=(n-1)f^{\ }_1(C)+f^{\ }_1(C')\,,\quad f^{\ }_N(C)=Nf_1(C)
\end{equation}
is a manifestation of the ``self-averaging'' of the logarithm (not \emph{any}
logarithm though, but this particular one), and we get
\begin{equation}
  \label{eq:3rdex''}
  \aver{\log\frac{2}{\cosh(2\theta_{\mathrm{tot}})+1}}=f^{\ }_N(C)
  =N\log\tau_1^{\ }\,.
\end{equation}
This is the Berry--Klein result (\ref{eq:logaver}), of course.

Finally, we turn to calculating $\aver{\tau^{\ }_N}$.
The first few $f_n^{\ }(C')$s are
\begin{eqnarray}
  \label{eq:firstfew}
f_1(C')&=&\frac{2}{C'+1}\,,\nonumber\\
f_2(C')&=&\frac{2}{C'+C}\,,\nonumber\\
f_3(C')&=&\frac{2}{\sqrt{(C'+1)(2C^2+C'-1)}}\,,
\end{eqnarray}
giving
\begin{eqnarray}
  \label{eq:firstavers}
 && \aver{\tau_2^{\ }}=f_2^{\ }(C)=\frac{1}{C}
    =\frac{\tau_1^{\ }}{2-\tau_1^{\ }}\,,\nonumber\\
 && \aver{\tau_3^{\ }}=f_3^{\ }(C)=\frac{2}{(C+1)\sqrt{2C-1}}
    =\frac{\tau_1^{\ }}{\sqrt{4/\tau_1^{\ }-3}}\,,
\end{eqnarray}
and it is frustratingly difficult to go beyond $n=3$.
But it is possible to evaluate the recurrence relation (\ref{eq:recurrence})
numerically and so determine $\aver{\tau^{\ }_N}=f^{\ }_N(C)$.
In passing, we note that 
$\aver{\tau_2^{\ }}^{\ }_{\mathrm{ray}}=\aver{\tau_2^{\ }}$ and
$\aver{\tau_3^{\ }}^{\ }_{\mathrm{ray}}>\aver{\tau_3^{\ }}$ for
${0<\tau_1^{\ }<1}$; ray optics fails for ${N>2}$.

\begin{figure}[t]
\begin{center}
\includegraphics[bb=120 466 437 692]{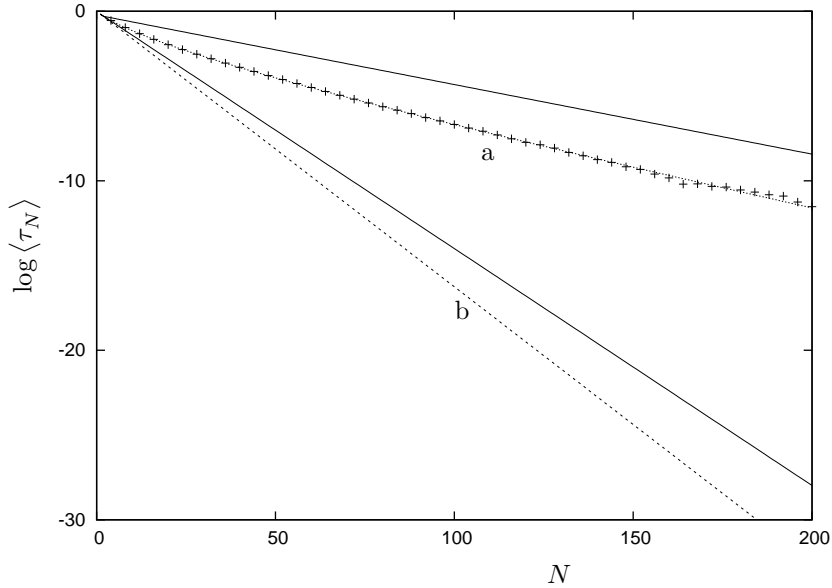}
\caption{\label{fig:logplot}%
    Average transmission probability for a stack of $N$ identical slabs.
    For $\tau_1^{\ }=0.85$ and ${N\leq200}$, the dotted curve `a' shows the
    values of $\log\aver{\tau_N^{\ }}$ computed by a numerical evaluation of
    the recurrence relation (\ref{eq:recurrence}), commencing with the
    small-$n$ functions of (\ref{eq:firstfew}). 
    The crosses that follow curve `a' are values obtained by a Monte Carlo
    calculation that simulated $400,000$ experimental realizations.
    The two solid lines are the upper and lower bounds of (\ref{eq:upperbound})
    and (\ref{eq:lower4}), respectively. 
    The dashed line `b' is the lower bound (\ref{eq:BK-lowlim}) derived by
    Berry and Klein~\cite{Berry+1:97}.
}
\end{center}
\end{figure}

For $\tau_1^{\ }=0.85$, the outcome of such a computation is shown in the
lin-log plot of Fig.~\ref{fig:logplot} as the dotted curve `a'.
The crosses near the curve were obtained by a Monte Carlo calculation in
which $400,000$ experiments were simulated with up to $200$ slabs.
The straight dashed line `b' is the lower bound of (\ref{eq:BK-lowlim}).
The solid lines are the upper and lower bounds discussed in the next section.
Other values of $\tau_1^{\ }$ result in plots with the same general features.

\section{Many slabs: Upper and lower bounds}
\label{sec:bounds}
\begin{figure}[t]
\begin{center}
\includegraphics{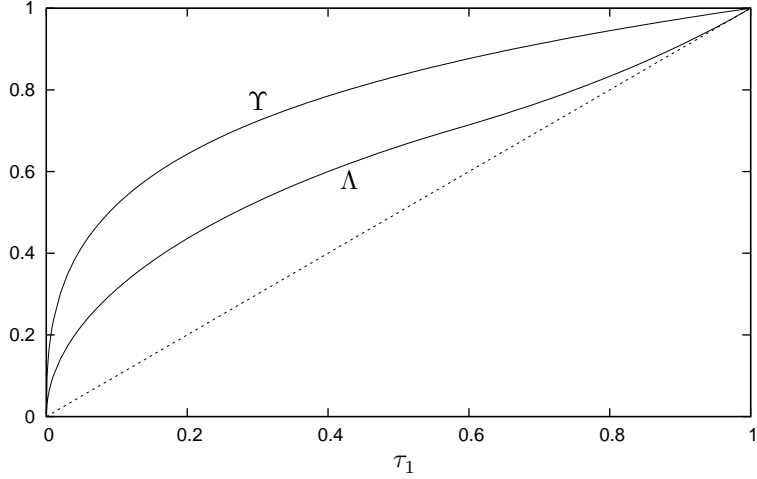}
\caption{\label{fig:bounds}%
   Upper bound $\Upsilon(\tau_1^{\ })$ and lower bound
   $\Lambda(\tau_1^{\ })$ on
   $({\smaver{\tau^{\ }_N}}/{\smaver{\tau^{\ }_2}})^{1/(N-2)}$
   as functions of $\tau_1^{\ }$.
   The dashed straight line is the lower bound on $\smaver{\tau^{\ }_N}^{1/N}$
   of (\ref{eq:BK-lowlim}).}
\end{center}
\end{figure}

Since $f_2^{\ }(C')\leq1/\sqrt{C'C}$, we have
\begin{eqnarray}
  \label{eq:upper1}
  f_3^{\ }(C')&\leq&
   \frac{1}{\sqrt{C'C}}\int\limits_{(2\pi)}\frac{\D\varphi}{2\pi}\,
            \frac{1}{\sqrt{C+S\,(S'/C')\cos\varphi}}\nonumber\\
&\leq&\frac{1}{\sqrt{C'C}}\int\limits_{(2\pi)}\frac{\D\varphi}{2\pi}\,
            \frac{1}{\sqrt{C+S\cos\varphi}}
\nonumber\\&=&\frac{1}{\sqrt{C'C}}\Upsilon(\tau_1^{\ })\,,
\end{eqnarray}
where the second inequality recognizes that the integral in the first line is a
monotonically increasing function of $S'/C'$, so that 
the replacement $S'/C'\to1$ increases its value.
The integral defining $\Upsilon(\tau_1^{\ })$ is of elliptic kind and its
value is less than $1$ if $C>1$, 
that is: $\Upsilon(\tau_1^{\ })<1$ if $\tau^{\ }_1<1$.
We conclude by induction that
\begin{equation}
  \label{eq:upper2}
  f_n^{\ }(C')\leq\frac{1}{\sqrt{C'C}}\Upsilon(\tau_1^{\ })^{n-2}
\end{equation}
holds for $n\geq2$. 
The upper bound
\begin{equation}
  \label{eq:upperbound}
  \aver{\tau^{\ }_N}\leq\aver{\tau^{\ }_2}\Upsilon(\tau_1^{\ })^{N-2}
\end{equation}
then follows.
The ray-optics result (\ref{eq:rayoptics}) is inconsistent with this upper
bound. 
Figure~\ref{fig:bounds} shows $\Upsilon(\tau_1^{\ })$ as a 
function of $\tau_1^{\ }$.

We derive a lower bound by first observing that
\begin{equation}
  \label{eq:lower1}
  f^{\ }_3(C')\geq f^{\ }_2(C')\Lambda(\tau^{\ }_1)
\end{equation}
with
\begin{equation}
  \label{eq:lower2}
  \Lambda(\tau^{\ }_1)=\min_{C'}\frac{ f^{\ }_3(C')}{f^{\ }_2(C')}
   =\left\{
    \begin{array}{c@{\ \mbox{for}\ }l}\displaystyle
      \frac{1}{2-\tau^{\ }_1} & \tau_1\geq2-\sqrt{2}\\[3ex] 
      \sqrt{\tau_1^{\ }-\tau_1^2/4}  & \tau_1^{\ }\leq2-\sqrt{2}
    \end{array}\right.
\end{equation}
and then inferring by induction that 
\begin{equation}
  \label{eq:lower3}
  f^{\ }_n(C')\geq f^{\ }_2(C') \Lambda(\tau^{\ }_1)^{n-2}
\end{equation}
holds for $n\geq2$.
The lower bound
\begin{equation}
  \label{eq:lower4}
  \aver{\tau^{\ }_N}\geq\aver{\tau^{\ }_2} \Lambda(\tau^{\ }_1)^{N-2}
\end{equation}
then follows. 
The plot of $\Lambda(\tau_1^{\ })$ as a function of $\tau_1^{\ }$ in
Fig.~\ref{fig:bounds} shows that ${\Lambda(\tau^{\ }_1)>\tau^{\ }_1}$ 
for ${0<\tau^{\ }_1<1}$ and, therefore, this lower
bound is more stringent than (\ref{eq:BK-lowlim}), but it is not tight either.
We are certain, however, that $\aver{\tau^{\ }_N}$ is bounded exponentially
both from above and from below.

\begin{figure}[t]
\begin{center}
\includegraphics[bb=110 466 437 692]{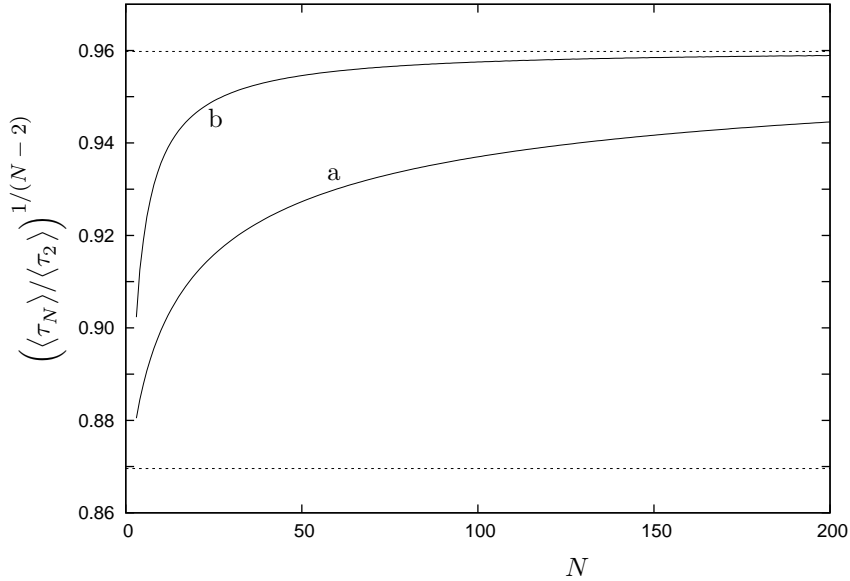}
\caption{\label{fig:linplot}%
    Values of $({\smaver{\tau^{\ }_N}}/{\smaver{\tau^{\ }_2}})^{1/(N-2)}$
    for ${\tau^{\ }_1=0.85}$.
    The bounds of (\ref{eq:bounds}) are the two horizontal dashed lines.
    Curve~`a' shows the actual values.
    The extrapolation explained in the context of (\ref{eq:conject}) and
    (\ref{eq:extrapol}) gives curve~`b'.
}
\end{center}
\end{figure}

Figure~\ref{fig:linplot} illustrates the two bounds
\begin{equation}
  \label{eq:bounds}
  \Lambda(\tau^{\ }_1)\leq
  \Biggl(\frac{\aver{\tau^{\ }_N}}{\aver{\tau^{\ }_2}}\Biggr)^{1/(N-2)}
  \leq \Upsilon(\tau^{\ }_1)
\end{equation}
for ${\tau^{\ }_1=0.85}$.
The values for curve~`a' are obtained by the numerical
evaluation of the recurrence relation (\ref{eq:recurrence}).
Clearly all values are well within the two bounds, the horizontal dashed lines.
This figure, and analogous plots for other values of~${\tau^{\ }_1}$, suggest
that 
\begin{equation}
  \label{eq:conject}
  \Biggl(\frac{\aver{\tau^{\ }_N}}{\aver{\tau^{\ }_2}}\Biggr)^{1/(N-2)}
  \longrightarrow  \Upsilon(\tau^{\ }_1)
\quad\mbox{as $N\to\infty$}\,.
\end{equation}
The corresponding observation in Fig.~\ref{fig:logplot} is that, for
sufficiently large $N$, line `a' there is parallel to the solid line for the
upper bound. 
At present, (\ref{eq:conject}) is no more than a conjecture that is supported
by a body of numerical evidence.

Some of the evidence is curve `b' in Fig.~\ref{fig:linplot}.
Its values are obtained by an extrapolation that assumes that 
\begin{equation}
  \label{eq:extrapol}
  \Biggl(\frac{\aver{\tau^{\ }_N}}{\aver{\tau^{\ }_2}}\Biggr)^{1/(N-2)}
  \approx A-B/N
\end{equation}
for large $N$ with $A$ and $B$ slowly varying with $N$. 
For two consecutive $N$ values of curve~`a' we can get an estimate of $A$ and
$B$, and curve~`b' represents the successive values of $A$ thus extrapolated. 
The rapid and consistent approach of `b' to the horizontal line of the upper
limit feeds the expectation that the conjecture (\ref{eq:conject}) could be
true. 
We leave the matter at that.

\section{Summary}
We established the recurrence relation (\ref{eq:recurrence}) that facilitates
the calculation of the average value $\aver{f(\tau_N^{\ })}$ of any function
of $\tau_N^{\ }$, the transmission probability through the stack of $N$
identical slabs with random gaps between them.
We observed that the individual values of $\tau_N^{\ }$ are spread over a
large range and, therefore, $\aver{\tau_N^{\ }}$ exceeds 
$\Exp{\smaver{\log\tau_N^{\ }}}=\tau_1^N$ by much.

Further, we derived strict upper and lower bounds on  $\aver{\tau_N^{\ }}$,
both bounds being exponential functions of $N$.
The ray-optics prediction for  $\aver{\tau_N^{\ }}$ is consistent with the
lower bound but not with the upper bound.
The upper bound, when used as an approximation for $\aver{\tau_N^{\ }}$, is of
much better accuracy than its derivation suggests and, based on numerical
evidence, we conjecture that it is asymptotically exact.

\section*{Acknowledgments}
We are grateful for discussions with Dominique Delande.
Centre for Quantum Technologies is a Research Centre of Excellence funded by
Ministry of Education and National Research Foundation of Singapore.

\end{document}